\documentclass[a4paper]{jpconf}
\usepackage{amssymb,amsfonts,amsmath}
\usepackage{graphicx}
\usepackage{subfigure}
\usepackage{hyperref}
\usepackage[numbers,square,sort&compress]{natbib}

\begin{document}
\title{Understanding the importance of transient resonances in extreme mass ratio inspirals}

\author{C~P~L~Berry$^{1,2}$, R~H~Cole$^{2}$, P~Ca\~{n}izares$^{3,2}$ and J~R~Gair$^{4,2}$}

\address{$^1$ School of Physics and Astronomy, University of Birmingham, Birmingham, B15 2TT, UK}
\address{$^2$ Institute of Astronomy, Madingley Road, Cambridge, CB3 0HA, UK}
\address{$^3$ Institute of Mathematics, Astrophysics and Particle Physics, Radboud University, Heyendaalseweg 135, 6525 AJ Nijmegen, Netherlands}
\address{$^4$ School of Mathematics, University of Edinburgh, Peter Guthrie Tait Road, Edinburgh EH9 3FD, UK}

\ead{cplb@star.sr.bham.ac.uk}

\begin{abstract}
Extreme mass ratio inspirals (EMRIs) occur when a compact object orbits a much larger one, like a solar-mass black hole around a supermassive black hole. The orbit has $3$ frequencies which evolve through the inspiral. If the orbital radial frequency and polar frequency become commensurate, the system passes through a \emph{transient resonance}. Evolving through resonance causes a jump in the evolution of the orbital parameters. We study these jumps and their impact on EMRI gravitational-wave detection. Jumps are smaller for lower eccentricity orbits; since most EMRIs have small eccentricities when passing through resonances, we expect that the impact on detection will be small. Neglecting the effects of transient resonances leads to a loss of $\sim4\%$ of detectable signals for an astrophysically motivated population of EMRIs.
\end{abstract}


The first observations of gravitational waves came from comparable mass stellar-mass binary black holes~\cite{Abbott2016d}. The evolving Laser Interferometer Space Antenna (eLISA) provides the chance to observe gravitational waves from stellar-mass black holes orbiting supermassive black holes~\cite{Amaro-Seoane2007}. These extreme mass ratio inspirals (EMRIs) emit $\sim10^4$--$10^5$ gravitational-wave cycles in eLISA's frequency band, allowing exquisite measurements to be made~\cite{Barack2004}. To detect and analyse EMRIs accurate waveform templates are required.

Flanagan and Hinderer~\cite{Flanagan2012} highlighted a previously overlooked phenomenon that complicates EMRI waveform modelling, that of transient resonances. During an inspiral, a resonance occurs when the radial frequency $\Omega_r$ and polar frequency $\Omega_\theta$ are a rational ratio of each other. Then, the orbit does not cover the whole allowed $r$--$\theta$ plane, but cycles over a single loop~\cite{Grossman2012}. The orbital phase can be expressed as
\begin{equation}
\varphi {} \sim {} \left(n_r \Omega_r - n_\theta \Omega_\theta\right) t + \left(n_r \dot{\Omega}_r - n_\theta \dot{\Omega}_\theta\right)t^2 + \ldots = \Omega t + 2\pi \frac{t^2}{\tau_\mathrm{res}^2} + \ldots
\end{equation}
On resonance $\Omega = 0$ so the first term, which normally dominates, goes to zero; the second term then governs dynamics for a duration set by the resonance time $\tau_\mathrm{res}$~\cite{Ruangsri2014,Berry2016a}. The evolution of the inspiral is determined by the gravitational self-force~\cite{Barack2009}; across a resonance, terms that usually average to zero can combine coherently, significantly impacting the orbital motion~\cite{Flanagan2012}.

For previous EMRI studies, adiabatic waveforms have been used. These average the self-force over the $r$--$\theta$ plane and do not include resonances. Adiabatic waveforms quickly dephase from true EMRI waveforms if passing through resonance causes a jump in the orbital parameters. The mismatch between adiabatic and the true waveforms could prevent signals from being detected or lead to biased parameter estimates.

In \cite{Berry2016a}, we investigated jumps in the orbital parameters from passing through a resonance. We used the same approximate self-force as Flanagan and Hinderer~\cite{Flanagan2012}. The size of jumps depends upon the magnitude of the self-force, the resonance time and the relative $r$--$\theta$ phase~\cite{Kevorkian1987,VanDeMeent2013,Berry2016a}. Jumps are most significant for low-order resonances, like the $2$:$3$ resonance, because higher-order resonances come closer the covering the entire $r$--$\theta$ plane (like nonresonant orbits). The jump magnitude is sensitive to the orbital parameters on resonance, and decreases for lower eccentricity or lower inclination orbits~\cite{Flanagan2012a}: there can be no resonances for circular orbits.


To assess the impact of resonances on detectability, we studied an astrophysical population of EMRIs that could be detectable with eLISA, assuming a two-year mission~\cite{Amaro-Seoane2012a}. We considered $10 M_\odot$ black holes inspiralling into $10^4$--$10^7 M_\odot$ supermassive black holes with a uniform distribution of spins. The initial eccentricity distribution followed Hopman and Alexander~\cite{Hopman2005}. As gravitational-wave emission circularises orbits~\cite{Peters1964}, we found that when important low-order resonances are encountered the eccentricity is small. Accordingly, the resonant jumps are also small. The match with adiabatic waveforms are good enough to still detect EMRIs in most cases~\cite{Berry2016a}. Considering the distribution of signal-to-noise ratios $\rho$ for the EMRIs before accounting for resonances and after including the reduction from imperfect waveforms, we find that $\sim4\%$ of signals drop below the assumed detection threshold $\rho_\mathrm{thres} = 15$.



In summary, transient resonances are a generic feature of EMRIs. Passing through resonance causes a jump in the evolution: the jump depends upon the phase on resonance, its size depends upon the orbit and the resonance. Ignoring resonances leads to waveform mismatch, but jumps should be small for most EMRIs because of their eccentricity. Therefore, detectability is not significantly reduced. The effect of resonances on parameter estimation is still to be investigated.

\ack
This work was supported in part by STFC and the Cambridge Philosophical Society. This is LIGO document reference LIGO-P1600349.

\bibliography{Resonances.bib}


\end{document}